\documentclass{pasj00}

\draft

\begin{document}
\SetRunningHead{S. Kato}{Frequency Correlations of QPOs by Warp Model}
\Received{2006/00/00}%{yyyy/mm/dd}
\Accepted{2007/00/00}%{yyyy/mm/dd}

\title{Frequency Correlations of QPOs Based on  a Disk Oscillation 
        Model in Warped Disks}

%%% begin:list of authors
\author{Shoji \textsc{Kato}}
    %\thanks{Example: Present Address is xxxxxxxxxx}}
\affil{Department of Informatics, Nara Sangyo University, Ikoma-gun,
       Nara 636-8503}
\email{kato@io.nara-su.ac.jp, kato@kusastro.kyoto-u.ac.jp}
\and
%\and
%\author{C-Firstname {\sc C-Familyname}}
%\affil{C-Address of Institute}\email{ccccc@xxx.xxx.xx.xx}
%%% end:list of authors

%%% Please use the following style in case that sorting by 
%%% affilation is impossible. 
%
% \author{%
%   D-Firstname \textsc{D-Familyname}\altaffilmark{1}
%   E-Firstname \textsc{E-Familyname}\altaffilmark{1,2}
% and
%   F-Firstname \textsc{F-Familyname}\altaffilmark{2}}
% \altaffiltext{1}{Address of Institute}
% \email{ddddd@xxx.xxx.xx.xx}
% \email{eeeee@xxx.xxx.xx.xx}
% \altaffiltext{2}{Address of Institute}

%% `\KeyWords{}' always has to be placed before `\maketitle'.
\KeyWords{accretion, accrection disks 
          --- quasi-periodic oscillations
          --- resonance
          --- neutron stars
          --- warp
          --- X-rays; stars} %Do NOT move this preamble from here!

\maketitle

\begin{abstract}
In previous papers we proposed a model that high-frequency quasi-periodic 
oscillations (QPOs) observed in black-hole and neutron-star X-ray binaries are
disk oscillations (inertial-acoustic and/or g-mode oscillations) resonantly excited 
on warped disks.
In this paper we examine whether time variations of  the QPOs and their 
frequency correlations observed in neutron-star X-ray binaries
can be accounted for by this disk-oscillation model.
By assuming that a warp has a time-dependent precession, we can well describe
observed frequency correlations among kHz QPOs and LF QPOs in a wide range of 
frequencies.
\end{abstract}

\section{Introduction}

Low-mass X-ray binaries (LMXBs) often show quasi-periodic oscillations
(QPOs) in their X-ray flux.
In the case of high-frequency QPOs (HF QPOs) in black-hole sources, they often 
appear in a pair and their frequencies are kept with time with the frequency 
ratio of 3 : 2.
In the case of kHz QPOs in neutron-star X-ray sources, they  
frequently appear also in pair, but with time variation. 
The frequency ratio is not kept to 3 : 2. 
However, the time variations of the pair kHz QPOs are not only correlated each other, 
but also correlated with the time variation of low-frequency QPOs (LF QPOs) (Boutloukos
et al. 2006).
The purpose of this paper is to examine whether these correlated time-variations
of QPOs in neutron-star X-ray binaries can be described as disk oscillations resonantly
excited on warped disks.

In disks deformed by some external forces,  excitation of
disk oscillations by resonant processes is generally expected.
A well-known example is the tidal instability in cataclysmic
variables (Whitehurst 1988; Hirose and Osaki 1990; Lubow 1991).
Another example is the excitation of spiral pattern in ram-pressure-deformed 
galactic disks (Tosa 1994; Kato and Tosa 1994).
In the context of high-frequency QPOs, a warp will be one of the most 
conceivable deformations of disks.
Based on this, we proposed a model that HF QPOs in balck-hole X-ray binaries and
kHz QPOs in neutron-star X-ray binaries  
are disk oscillations resonantly excited on disks deformed by warp
(e.g., Kato 2003, 2004a,b; Klu{\' z}niak et al. 2004; Kato 2005a,b; Kato and Fukue 2006).

In this warp model there are three possible combinations of i) type of resonance
and ii) type of oscillations.
Stability analyses for these three cases show that inertial-acoustic
oscillations and/or g-mode oscillations are excited by their horizontal coupling
with a warped disks (Kato 2004b). 
An overeview of the resonant non-linear coupling processes among oscillations
and warp, which feedback to the original oscillations so as to amplify or dampen 
the oscillations, are shown in figure 1 of Kato (2004b).
By this resonant model in warped disks, we can account for the 3 : 2 frequency
ratio of HF QPOs in black-hole X-ray sources (e.g., Kato 2004b; Kato and Fukue 2006).

Although the above analyses of stability by Kato (2004b) is only to the case in which
the warp has no precession, we think that the excitation of inertial-acoustic oscillations 
and/or g-mode oscillations still holds even in the case in which the warp has precession.
We further assume that the warp has time-dependent precession in the 
case of neutron stars.\footnote{
In the case of black-hole sources we assume that the warp has no precession.
This difference will be related to the difference of the surface of the central sources.
If the surface is present, magnetic couplings and radiative couplings (e.g., Pringle 1996; 
Maloney et al. 1996) of the disks with the central sources will cause precession of warps.
}
By the time-dependent precession, the resonant radius changes with time.
Hence, frequencies of resonant oscillations vary with time.
During the frequency changes of resonant oscillations, they  
are correlated each other in this warp model as shown below, since the resonance 
occurs at the same radius for different oscillation modes. 

In this paper we demonstrate that the observed correlations of kHz QPOs and 
LF QPOs  in neutron-star X-ray binaries can be described by this 
warp model with time-dependent precession. 
%Comparison of QPO frequencies by our warped-disk model with observations
%is done on the diagram drawn by Bursa on lower QPO kHz frequency versus 
%upper kHz QPO frequency (e.g., Abramowicz 2006), and on figures obtained 
%to Cir X-ray by Boutloukos and van der Klis (2006).  

\section{Overview of the Resonant Oscillation Model in Warped Disks}

We present here the main part of the warp model, although the model has various 
variations.
Let us consider disk oscillations described by ($\omega$, $m$, $n$), 
where $\omega$ and $m$ are angular frequency and azimuthal wavenumber ($m=0,1,2,...$) 
of the oscillations, respectively, and $n$ is an integer ($n=0,1,2,...$) describing node 
number of the oscillations in the vertical direction (e.g., see Kato et al. 1998; Kato 2001).
For a set of ($\omega$, $m$, $n$), there are two different kinds of
modes of oscillations, except for the case of $n=0$.
In the case of $n=0$, we have inertial-acoustic oscillations alone, while
in the case of $n\geq 1$ we have two different modes of oscillations. 
One is gravity mode, and the other is corrugation mode ($n=1$) or vertical-acoustic 
mode ($n\geq 2$) (see Kato et al. 1998; Kato 2001).

Now, the disks are assumed to be warped with a time-dependent precession. 
The warp is a kind of a global one-armed corrugation waves, and is described 
by ($\omega_{\rm p}$, 1, 1), where $\omega_{\rm p}$ is the angular frequency
of the precession, $\omega_{\rm p}>0$ being prograde and  $\omega_{\rm p}<0$ 
retrograde.

\subsection{Resonant condition and resonant radius}

Nonlinear resonant interaction between a warp with ($\omega_{\rm p}$,  $1$, $1$) and 
an oscillation with ($\omega$, $m$, $n$) brings about
oscillations described by ($\omega\pm \omega_{\rm p}$, $m\pm 1$, $n\pm 1$),
where arbitrary combinations of $\pm$ are possible.
(These oscillations are called hereafter intermediate oscillations.)
These intermediate oscillations have resonant interaction with the disk at 
particular radii where the dispersion relation of the intermediate oscillations
is satisfied.
There are two kinds of resonance, corresponding to the fact that two kinds 
of oscillation modes are described by the same dispersion relation, i.e., 
i) inertial-acoustic oscillations and gravity oscillations, and 
ii) vertical-acoustic oscillations.
We call the resonance related to the former oscillations horizontal resonance,
while the resonance related to the latter oscillations vertical resonance
(e.g., see Kato 2004b).

After making the resonant interaction with the disk, the intermediate oscillations
nonlinearly couple with the warp to feedback to the original oscillations 
($\omega$, $m$, $n$) (see figure 1 of Kato 2004b).
This nonlinear feedback processes amplify or dampen the original oscillations, 
since a resonance process is involved in the feedback processes. 
Careful stability analyses which resonance excites oscillations and which oscillations
are excited have been made in the case of no precession (Kato 2004b).
The results show that inertial-acoustic oscillations and/or gravity oscillations are 
excited by horizontal resonance (Kato 2004b).
This result will not change even when there is precession.
Hence, we hereafter restrict our attention only on the above case,
i.e., inertial-acoustic oscillations and/or gravity oscillations which have 
resonant interaction with the warped disk through the horizontal resonance.

For the resonance to occur effectively, the place of resonance and
the place where the oscillations predominantly exist must be the same.
We find that in the case of no precession, the condition is realized at the radius  
of $\kappa=\Omega/2$, 
where $\kappa$ is the epicyclic
frequency and $\Omega$ is the angular velocity of disk rotation.
This can be simply extended to the case where the warp has precession, which gives
the resonant condition as [see also Kato (2005a)\footnote
{In Kato (2005), $\omega_{\rm p}>0$ was retrograde, but here we adopt 
$\omega_{\rm p}>$ for prograde precession.}] 
\begin{equation}
   \kappa={1\over 2}(\Omega + \omega_{\rm p}).
\label{res}
\end{equation}
Hereafter, $\Omega$ is taken to be the Keplerian angular velocity, $\Omega_{\rm K}$,
when numerical figures are necessary.
Equation (\ref{res}) gives the resonant radius, $r_{\rm r}$,  as a function
of $\omega_{\rm p}$, when the mass and spin of central source are given.
The $r_{\rm r}$--$\omega_{\rm p}$ relation is shown in figure 1 for 
the case in which the mass of
the central source is $2M_\odot$ and the central source has no spin
(i.e., the metric is the Schwarzschild one).
It is noted that when the mass of the central source is smaller than $2M_\odot$ by
a factoe $\alpha$, the same $r_{\rm r}/r_{\rm g}$ 
is realized for the precession faster than that in the case of $2M_\odot$ by factor 
$\alpha^{-1}$.  

%---------------------------------------------------------------------------------------
\begin{figure}
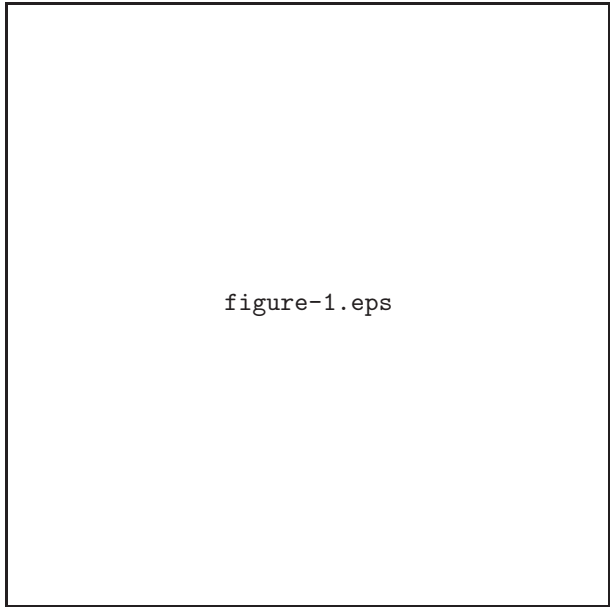

  \begin{center}
    %\FigureFile(80mm,80mm){precession-radius-02-meshlow.eps}
    \FigureFile(80mm,80mm){figure-1.eps}
    %%% \FigureFile(width,height){filename}
  \end{center}
  \caption{The resonant radius versus frequency of precession.
           The mass of the central source is $2M_\odot$.
           The metric is taken to be the Schwarzschild, i.e., $a_*=0$.
           In disks in which the warp has no precession, the resonance occus at $4r_{\rm g}$.
           When the precession is prograde ($\omega_{\rm p}>0$), the resonant radius 
           becomes smaller than $4r_{\rm g}$ as the precession frequency increases, 
           while becomes larger when the precession is retrograde ($\omega_{\rm p}<0$)
           and it absolute value increases.
           In the case of retrograde precession, however, the resonant radii are not unique 
           for a given $\omega_{\rm p}$, as shown in the figure. 
           That is, another resonance (resonance at an outer radius) appears
           far outside of the disk.
           The radius of the outer resonance 
           moves inwards, as the absolute value of precession increases.
           The inner and outer resonances join together for a certain value of
           retrograde precession, and no resonance occurs for a larger value of retrograde
           precession.}
%\label{fig:figure 1}
\end{figure}
%----------------------------------------------------------------------------------------

The resonant condition in the case of no precession, $\kappa=\Omega/2$,
is realized at $4r_{\rm g}$ when there is no spin.
Figure 1 show that if the precession is prograde, the resonant 
radius becomes smaller than $4r_{\rm g}$ with increase of precession frequency,
 $\omega_{\rm p}$.
If the precession is retrograde, however, we have resonance at two different radii.
One is close to $4r_{\rm g}$ when the precession is small, and moves outward 
with increase of the absolute value of $\omega_{\rm p}$.
The other one is far outside of the disk when the precession is weak,
and the radius of resonance moves inward with increase of the absolute value of precession. 
At a certain value of retrograde precession, both radii coincide and
no resonance occurs for retrograde precession with faster precession.

\subsection{frequencies of resonant oscillations}

Since inertial-acoustic oscillations or g-mode oscillations are concerned here,
the place where the oscillations exist predominantly is the place where  
$(\omega-m\Omega)^2-\kappa^2\sim 0$ is satisfied (e.g., see Kato and
Fukue 2006).
In other words, frequencies of resonant oscillations are $(m\Omega\pm \kappa)_{\rm r}$,
where various $m$'s are allowed, and the subscript r denotes the values at the resonant radius.
The axisymmetric oscillations with $m=0$ will be observationally less interesting by the
nature of symmetry itself.
Hence we consider only non-axisymmetic oscillations.
Typical non-axisymmetric modes of the oscillations are those with 
$m=1$ and $m=2$.
Hence, as the frequencies of such oscillations, we have
$(\Omega-\kappa)_{\rm r}$ (i.e., $m=1$), $(2\Omega-\kappa)_{\rm r}$, (i.e., $m=2$), 
$(\Omega+\kappa)_{\rm r}$ (i.e., $m=1$) ,$(2\Omega+\kappa)_{\rm r}$ (i.e., $m=2$),...
For convenience, we introduce the following notations defined by
\begin{equation}
   \omega_{\rm LL}=(\Omega-\kappa)_{\rm r}, \quad
   \omega_{\rm L}=(2\Omega-\kappa)_{\rm r}, \quad
   \omega_{\rm H}=(\Omega+\kappa)_{\rm r}.
\label{2}
\end{equation}

Here, the relation between frequencies of resonant oscillations and of 
the observed QPOs should be briefly discussed. 
In the case of black-hole X-ray binaries, we think that the one-armed oscillations with 
frequency $\omega_{\rm LL}$ will be observed with the two-fold frequency, 
$2\omega_{\rm LL}$, in addition to $\omega_{\rm LL}$ itself 
by the following reasons (Kato and Fukue 2006).
In black-hole sources, the high-frequency QPOs are observed only in the phase where 
the sources are at the steep power-law state (i.e., very high state)
(Remillard 2005).
In such states, the disk region where the QPOs are excited  
will be inside a compact hot torus, and the observed QPO photons are those which are
Comptonized in the torus.
In such cases, observed Comptonized high energy photon from  
one-armed oscillations ($m=1$) will have two maxima during one 
cycle of the oscillations (see figures 2 -- 4 of Kato and Fukue 2006).
This means that $2\omega_{\rm LL}$ will be observed with large amplitude (in many
cases with an amplitude larger than that of the oscillation of $\omega_{\rm LL}$), 
since the oscillation is one-armed (Kato and Fukue 2006). 
Based on this consideration, we have suggested that the observed pair QPO
frequencies in black hole sources are $\omega_{\rm L}$ and $2\omega_{\rm LL}$ 
(not $\omega_{\rm LL}$) (see, e.g., Kato 2004b; Kato and Fukue 2006).

Even in the case of neutron-star X-ray binaries, a similar situation may exist.
Including this possibility, we regard $\omega_{\rm H}$,  $\omega_{\rm L}$,
$2\omega_{\rm LL}$,  and $\omega_{\rm LL}$ as the main candidates of observed 
QPO frequencies in neutron stars. 

\subsection{Frequency-Frequency Relations}

Frequencies $\omega_{\rm H}$, $\omega_{\rm L}$, $2\omega_{\rm LL}$,
$\omega_{\rm LL}$, and $\omega_{\rm p}$ 
are given functions of the resonant radius $r_{\rm r}$, 
spin parameter $a_*$, and the mass $M$ of the central source.
Hence, eliminating $r_{\rm r}$ from these expressions for the frequencies, we obtain
relations among $\omega_{\rm H}$, $\omega_{\rm L}$, $2\omega_{\rm LL}$, 
$\omega_{\rm LL}$, and $\omega_{\rm p}$.
Parameters are $a_*$ and $M$.
Figure 2 shows the relations by taking $\omega_{\rm L}$ as the abscissa in the case of
$a_*=0$ and $M=2M_\odot$.
The straight line of $\omega_{\rm L}$ -- $\omega_{\rm L}$ relation is also shown for
comparison.
This figure should be compared with figure 2.9 of van der Klis (2004).
The latter shows observed frequency-correlations among various QPOs in neutron-star LMXBs.
Comparison suggests that the observed frequency correlations of QPOs seem to be
qualitatively described by the present disk oscillation model. 
More close comparison between observations and the model is made in the next section.

%-----------------------------------------------------------
\begin{figure}
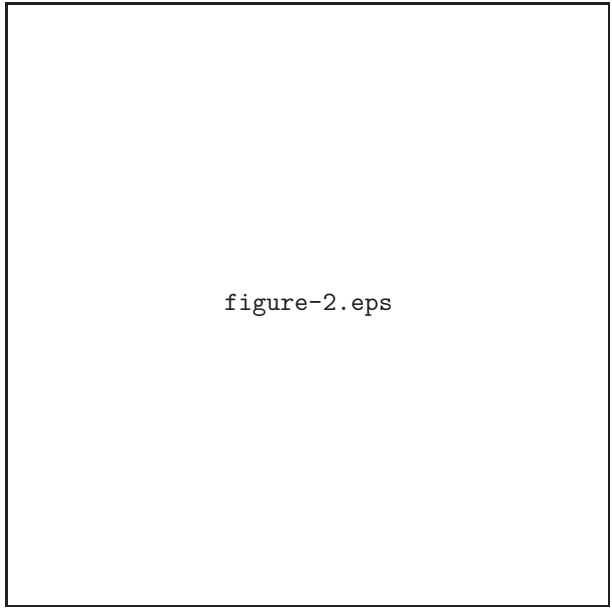

  \begin{center}
    \FigureFile(80mm,80mm){figure-2.eps}
    %%% \FigureFile(width,height){filename}
  \end{center}
 \caption{Various frequencies as functions of $\omega_{\rm L}$.
The frequency of precession shown here, $\omega_{\rm p}$, is the absolute value.
In the main part of this figure, the value of $\omega_{\rm p}$ is negative (retrograde).
The values of parameters adopted here are $M=2M_\odot$ and  $a_*=0$
} 
%\label{fig:figure 2}
\end{figure}
%-------------------------------------------------------------------------------------------

\section{Correlation of QPO Frequencies}

\subsection{Hectohertz QPOs}

In atoll sources (less luminous neutron-star LMXBs), hectohertz QPOs
(hHz QPOs) have been observed (see figure 2.9 of van der Klis 2004).
Their frequencies are in the range of 100 -- 200 Hz, and seen in atoll sources in 
most state.
Their presence in Z sources is, however, uncertain.
Different from kHz QPOs, their frequencies are approximately constant, which
is similar across sources (van der Klis 2004).

The hectohertz QPOs can be intrerpreted in the present model as observations 
of warp.
Figure 2 shows that for a wide range of variations of $\omega_{\rm L}$,
$\omega_{\rm p}$ is approximately constant around 100 -- 200 Hz, which is
consistent with the observational characteristics of hHz QPOs.
The fact that the value of $\vert\omega_{\rm p}\vert$ remains around 
100 -- 200 Hz is related to the fact that the lower limit of $\omega_{\rm P}(<0)$ shown 
in figure 1 is around 100 -- 200 Hz.
The lower limit of $\omega_{\rm P}(<0)$ depends on $a_*$.
If $a_*>0$, the maximum value of $\vert \omega_{\rm p}\vert$ slightly decreases
from that in the case of $a_*=0$, while it increases with decrease of the mass of 
the central source.

 \subsection{Correlation of pair kHz QPOs}

As argued in  previous papers (e.g., Kato 2004b; Kato and Fukue 2006),  
we think that the high-frequency 
pair QPOs in black-hole LMXBs are oscillations of $\omega_{\rm L}$ and 
$2\omega_{\rm LL}$.
Their frequency ratio is just 3 : 2, since we assume that in black-hole sources
the warp has no precession.\footnote{
The reason why there is a difference of disk precession in black-hole and neutron-star
sources is a subject to be clarified, but we suppose that it will be related to the 
difference of surfaces of the central sources. }
In the case of neutron-star LMXBs, we assume that the warp has time-dependent precession 
although the pair oscillations are still $\omega_{\rm L}$ and $2\omega_{\rm LL}$.
We examine here whether the frequency correlation in kHz QPOs can be accounted for
by this picture.
 
Figure 3 shows $\omega_{\rm L}$ as a function of
$2\omega_{\rm LL}$ for $a_*=0$ in the frequency range in which the 
ratio of the above two frequencies is around 3 : 2, i.e.,
the precession is not too large.
For comparison, the $\omega_{\rm H}$ -- $2\omega_{\rm LL}$ relation is also shown. 
In figure 3,  the mass $M$ of the central sources is taken so that 
$2\omega_{\rm LL}$ becomes 600Hz in the case of no precession.
This means that we have adopted $M=2.4M_\odot$, since 
$2\omega_{\rm LL}$ is given by
$2\omega_{\rm LL}=1.43\times 10^3(M/M_\odot)^{-1}$ Hz  
in the case of $a_*=0$ (e.g., see Kato and Fukue 2006).

Bursa (2003, see also Abramowicz 2005, Klu{\'z}niak 2005)
plotted the observed data of pair QPOs of some typical neutron-star sources 
on a diagram of the upper kHz QPO frequency
versus lower kHz QPO frequency, in order to see how their time
changes are correlated.
The plots obtained by Bursa have been superposed on figure 3 by regarding 
the lower QPO frequencies as $2\omega_{\rm LL}$.

Figure 3 shows that the observed correlated changes of the upper and 
lower QPO frequencies are qualitatively described by
the correlated changes of $\omega_{\rm L}$  (or $\omega_{\rm H}$) and
$2\omega_{\rm LL}$.

%----------------------------------------------------------------------------
\begin{figure}
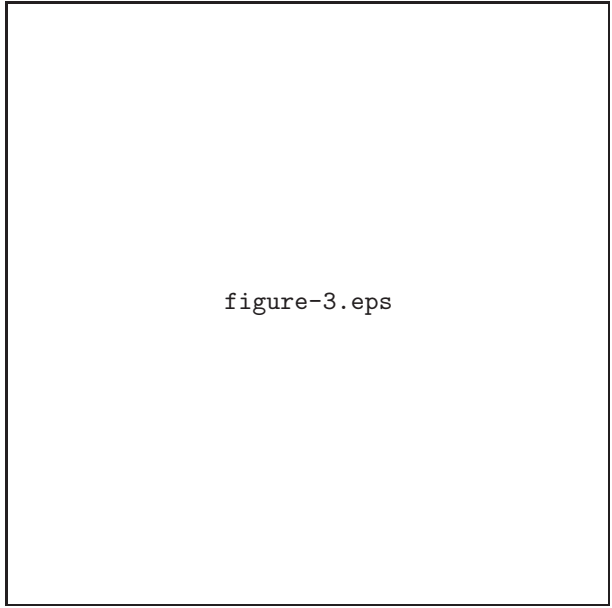

  \begin{center}
    \FigureFile(80mm,80mm){figure-3.eps}
    %%% \FigureFile(width,height){filename}
 \end{center}
  \caption{
  Dependence of $\omega_{\rm H}$ and $\omega_{\rm L}$ on $2\omega_{\rm LL}$
  in the case of $a_*=0$ and $M=2.4M_\odot$.
  The plots of the upper versus lower frequencies of the observed pair kHz QPOs 
  obtained by Bursa (2003) are superposed by taking the frequency of the lower kHz QPOs
  as $2\omega_{\rm LL}$.
  This figure shows that the plots of the observed data are between two curves of
  $\omega_{\rm H}$ -- $2\omega_{\rm LL}$ and 
  $\omega_{\rm L}$ -- $2\omega_{\rm LL}$.
  Bursa's diagram is taken from Abramowicz (2005).  
}
  \label{fig:3}
 % \end{center}
\end{figure}
%---------------------------------------------------------------------------------------

\subsection{KHz QPOs in low frequency range and LF QPOs}

As shown in figure 1, resonance occurs at two different radii when precession is
retrograde.
One is close  to $4r_{\rm g}$, while the other is at an outer region.
In some sources only the oscillations in the outer resonance will be observed,
since the inner region of disks may be highly perturbed  from the steady state 
by magnetic and/or radiative disturbances from the central source.
Here, we consider characteristics of resonant oscillations which are excited at the
outer resonant radius.
In order to compare them with observational results,
dependences of $\omega_{\rm H}$, $\omega_{\rm L}$, $\omega_{\rm p}$, 
and $2\omega_{\rm LL}$ on $\omega_{\rm LL}$ are shown in figure 4.
In addition to them, the $(3\Omega+\kappa)_{\rm r}$ -- $\omega_{\rm LL}$
relation is drawn in figure 4, where
$(3\Omega+\kappa)_{\rm r}$ is the frequency of one of resonant oscillations
with $m=3$.
There are other oscillation modes with high frequency.
Among such oscillations, the oscillations of $(3\Omega-\kappa)_{\rm r}$ (i.e., $m=3$) 
have frequencies close to $\omega_{\rm H}$, and thus it is not shown here.
Resonant oscillations with $(2\Omega+\kappa)_{\rm r}$ are also not shown in figure 4, 
since the curve of the $(2\Omega+\kappa)_{\rm r}$ -- $\omega_{\rm LL}$ relation is 
between two curves of  the $(3\Omega+\kappa)_{\rm r}$ -- 
$\omega_{\rm LL}$ and of the $\omega_{\rm H}$ -- $\omega_{\rm LL}$ relations.  
The straight line of $\omega_{\rm LL}$ -- $\omega_{\rm LL}$ is shown in the figure.

%-------------------------------------------------------------------------------
\begin{figure}
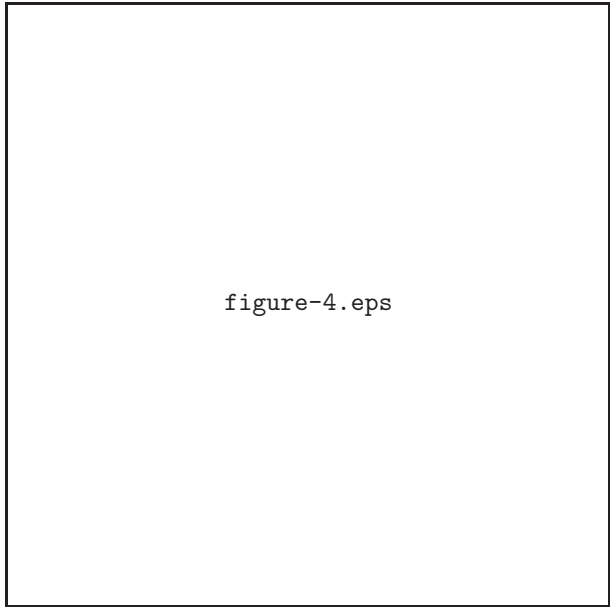

  \begin{center}
    \FigureFile(80mm,80mm){figure-4.eps}
    %%% \FigureFile(width,height){filename}
  \end{center}
  \caption{Dependences of various frequencies of resonant oscillations on 
$\omega_{\rm LL}$.
The frequencies considered here are in a low frequency range in order to compare 
them with those observed in Cir X-1. 
The most upper curve is the $(3\Omega+\kappa)_{\rm r}$ -- $\omega_{\rm LL}$
relation.
Figure 6 of Boutloukos et al. (2006), which shows the observed frequency -- frequency
correlations in Cir X-1, has been superposed assuming that $\nu_{\rm LF}$ corresponds
to $\omega_{\rm LL}$. 
} 
%\label{fig:figure 4}
\end{figure}
%-------------------------------------------------------------------------------

In order to examine whether the curves drawn in figure 4 can account for observational 
frequency-frequency correlations, the plots of observational data for Cir X-1 by  
Boutloukos et al. (2006) (figure 6 of their paper) are superposed on figure 4, assuming that 
the frequency of LF QPOs corresponds to $\omega_{\rm LL}$.
The superposed fugure seems to show that our disk oscillation model can well describe 
the observed frequency correlations.
That is, $\omega_{\rm LL}$ corresponds to the frequency of LF QPOs and 
$\omega_{\rm L}$ to that of the lower kHz QPOs.
The frequency $\omega_{\rm H}$, however, seems to be slightly lower than the observed 
upper frequency of kHz QPOs.
As frequencies of resonant oscillations whose frequencies are higher than $\omega_{\rm H}$, 
we have $(2\Omega+\kappa)_{\rm r}$, $(3\Omega+\kappa)_{\rm r}$,...
Among them, the frequency $(3\Omega+\kappa)_{\rm r}$ seems to well describe the 
observed data of the upper kHz QPOs.
The reason why oscillations of $(3\Omega+\kappa)_{\rm r}$ dominate over those of
$(2\Omega+\kappa)_{\rm r}$ and $\omega_{\rm H}$ is a subject to be discussed further.

When resonance occurs at an inner region of disks,  $2\omega_{\rm LL}$ is
interpreted as the frequency of the lower kHz QPOs, as shown in figure 3.
However, in the present case in which resonance occurs at an outer region,
$\omega_{\rm LL}$ (and $2\omega_{\rm LL}$)  is no longer the frequency of the lower 
kHz QPOs,  since it is too low.
It becomes the frequency of LF QPOs.
The reason why $\omega_{\rm LL}$, not $2\omega_{\rm LL}$, is the frequency of 
LF QPOs will be the followings.
The outer region of the disk will not be covered with a hot torus.
Thus, different from the case of black-hole X-ray sources, the resonant region will
be outside a hot torus.
In such case one-armed oscillations will be observed by the frequency of the oscillations
themselves, not by two-fold ones. 
The inertial-acoustic oscillations with $m=1$, however, propagate inward from
the resonant radius (see the propagation region shown in figure 6 of Kato and Fukue 2006).
Hence, if the oscillations propagate inward and enter into an hot torus, the oscillations
of two-fold frequency will be observed (see Kato and Fukue 2006).      
This may be one of possible causes of the occasional appearance of the two-fold 
frequencies of $\omega_{\rm LL}$ in Cir X-1.
 
%Figure 4 seems to suggest that the vertical spread of observational data of lower 
%kHz QPOs is due to mixture of three frequencies, $\omega_{\rm p}$, 
%$\omega_{\rm L}$, and $\omega_{\rm H}$. 
 
In summary, the LF QPOs will be manifestation of the oscillations of $\omega_{\rm LL}$,
and the lower kHz QPOs  ($\nu_{\ell}$ in notations of Boutloukos et al.
2006) will be a mixture of $\omega_{\rm p}$, $\omega_{\rm L}$, and 
$\omega_{\rm H}$.
The upper kHz QPOs ($\nu_u$ in notations of Boutloukos 2006) are suggested to be 
oscillations of higher $m$ modes such as  those with $(3\Omega+\kappa)_{\rm r}$ or 
$(2\Omega+\kappa)_{\rm r}$.

\section{Discussion}

In this paper we have suggested that the pair kHz QPOs and the low-frequency
QPOs in neutron-star LMXBs are qualitatively described, in a wide 
frequency range, as resonantly excited disk-oscillations in warped disks.
The oscillations are non-axisymmetric inertial-acoustic or g-mode oscillations 
of $m=1$, $m=2$,  and sometimes $m=3$. 
Both inertial-acoustic oscillations  and g-mode oscillations are possible
candidates of QPOs, but the former will be better, since the magnitudes of 
temperature and density variations associated with the oscillations are larger in the
former than in the latter.
Especially,  the oscillations of $\omega_{\rm LL}$ will be inertial-acoustic
oscillations, since they propagate inwards from the resonant radius and thus
observational appearance of the harmonics, $2\omega_{\rm LL}$, will be 
conceivable (see subsection 3.3).

In the present disk oscillation model, the cause of correlated frequency-changes 
of various QPOs is time change of resonant radius resulting from 
time-dependent precession of the warp.
Has the warp been observed?
We think that it has been observed as the hectoherz QPOs in atoll sources
(see figure 2, in which $\omega_{\rm p}$ is around 100 Hz $\sim$ 200 Hz
in a wide frequency range of $\omega_{\rm L}$).
Furthermore,  in sources in which kHz QPOs appear in low frequency region, 
precession of the warp will be a cause of vertical spread of $\nu_\ell$
in figure 4.
The closeness of $\omega_{\rm p}$ and $\omega_{\rm L}$ in low frequency region
comes from the following situation.
The resonant condition, $\kappa=(\Omega+\omega_{\rm p})/2$ gives
$\omega_{\rm p}=(2\kappa-\Omega)_{\rm r}$, which is $\sim \Omega_{\rm r}$ 
when the resonant radius is far from the innermost region since there 
$\kappa\sim\Omega$.
On the other hand, $\omega_{\rm L}=(2\Omega-\kappa)_{\rm r}\sim \Omega_{\rm r}$ 
in such a region.

It is emphasized that by difference of frequency range in consideration, 
the counterparts of the oscillation modes to observed QPOs are different.
That is,  in the case in which the frequencies of kHz QPOs are high (subsection 3.2),
the counterpart of the lower kHz QPOs is $\omega_{\rm LL}$ (more exactly,
$2\omega_{\rm LL}$) (see figure 3).
In the sources in which  the frequencies of kHz QPOs are low (subsection 3.3), however, the 
counterpart of the lower kHz QPOs is $\omega_{\rm L}$, not $\omega_{\rm LL}$.
In the latter case, the oscillation of $\omega_{\rm LL}$ corresponds to 
$\nu_{LF}$ (see figure 4).
 
There are some problems to be examined or to be clarified further.
First,  it is known that the frequencies of lower kHz QPOs, $\nu_\ell$,  and those of 
horizontal branch QPOs, $\nu_{\rm HBO}$, are correlated as
$\nu_\ell \sim 0.08 \nu_{\rm HBO}$ (Psaltis et al. 1999; Belloni et al. 2002).
In the framework of our present model, there are no oscillation modes corresponding to
the horizontal branch QPOs.
They might be related to resonant oscillations resulting from other types of
resonance.
Even in the framework of the resonant oscillation model in warped disks, there are other 
types of resonant osciilations (Kato 2004a; 2005b), although their excitation is uncertain.
%The low-frequency QPOs and the horizontal barnch QPOs seem to be somewhat
%different (see also Boutloukos et al. 2006). 
Second, a detailed inspection of figure 3 shows that the 
$\omega_{\rm L}$--$2\omega_{\rm LL}$ relation cannot always well
describe the observed data.
%For example, the relation can well account for the observed data of Sco X-1,
%as long as the frequencies are close to 3:2 ratio.
%However, the data seem to deviate from the 
%$\omega_{\rm L}$--$2\omega_{\rm LL}$ curve when the lower frequency is 
%as high as 800 Hz.
That is, to sources whose lower frequencies are lower than 600 Hz (i.e.,
GX 4340+0 and GX-5), the observed data are generally above the
$\omega_{\rm L}$--$2\omega_{\rm LL}$  curve on the diagram,  rather close to
the $\omega_{\rm H}$--$2\omega_{\rm LL}$ curve.
A preliminary study seems to show that it is difficult to explain the deviation by adjusting
the values of parameters (mass and spin of the central source). 
If we want to explain the difference by a deviation of disk rotation from the
Keplerian one,  a rather large deviation is necessary.
If $\omega_{\rm H}$ is taken as the upper QPO frequency below 900 Hz
and $\omega_{\rm L}$ above 900 Hz, qualitative agreement with observations
becomes better.
However, it seems not to be clear whether such choice of oscillation modes
is physically acceptable.
Some more consideration concerning the cause of the deviation is necessary, which is a 
subject in the future.

%conceivale causes are that (i) the disk rotation deviates from the Keplerian one,
%(ii) the frequency of resonant oscillations is changed by radial propagation of the 
%oscillations, or
%(iii) some non-linear effects affect so as to change the resonant radius and frequency.
%The deviation of the observational data from the $\omega_{\rm L}$ --
%$2\omega_{\rm LL}$ curve is a subject to be examined further in the future.
 
\bigskip
\leftskip=20pt
\parindent=-20pt
\par
{\bf References}
\par
%Abramowicz, M. A., \& Klu{\' z}niak, W. 2001, A\&A, 374, L19 \par
Abramowicz, M.A. 2005, Astron. Nachr. 326, No.9 \par
Belloni, T., Psaltis, D., van der Klis, M. 2002, ApJ., 572, 392 \par
Boutloukos, S., van der Klis, M., Altamirano, D., Klein-Wolt, M., Wijnands, R., 
      Jonker, P.G., Fender, R.P. 2006, astro-ph/0608089 \par
Bursa, M. 2003, unpublished \par 
Hirose, M., Osaki, Y. 1990, PASJ 42, 135\par
%Homan, J. Klein-Wolt, M., Rossi, S. Miller, J.M., Wijnands, R., Belloni, 
%       T., van der Klis, M., Lewin, W.H.G., 2003, ApJ, 586, 1262 \par
Kato, S. 2001, PASJ, 53, 1\par 
%Kato, S. 2003a, PASJ, 55, 257 \par
Kato, S. 2003, PASJ, 55, 801\par
Kato, S. 2004a, PASJ, 56, 559 \par
Kato, S. 2004b, PASJ, 56, 905\par
%Kato, S. 2004c, PASJ, 56, L25\par
Kato, S. 2005a, PASJ, 57, L17 \par
Kato, S. 2005b, PASJ, 57, 679 \par
Kato, S., Fukue, J., \& Mineshige, S. 1998, Black-Hole Accretion Disks 
  (Kyoto: Kyoto University Press)\par
Kato, S., Tosa, M. 1994, PASJ, 46, 559 \par
Kato, S., Fukue, J. 2006, PASJ , 58, 909\par
%Klu{\' z}niak, W., \& Abramowicz, M. 2001, Acta Phys. Pol. B32, 3605   \par
Klu{\' z}niak, W. 2005, Astron. Nachr. 326, 820 \par 
Klu{\' z}niak, W., Abramowicz, M. A., Kato, S., Lee, W. H., \& Stergioulas,
   N. 2004, ApJ, 603, L89 \par 
%Klu{\' z}niak, W., Lasota, J-P., \& Abramowicz, M.A. 2005, 
%      astro-ph 0503151\par
%Lamb, F. K. \& Miller, M. C. 2003, astro-ph/0308179 \par
%Li, L.-X., Goodman, J., Narayan, R. 2003, ApJ, 593,980 \par
Lubow, S.H. 1991, ApJ, 381, 259\par
%McClintock, J.E., Remillard, R.A. 2005, "Black Hole Binaries", in
%  Compact Stellar X-ray Sources, eds. W.H.G. Lewin and M. van der Klis,
%   Cambridge University Press, Cambridge, in press; astro-ph/0306213 \par
Psaltis, D., Belloni, T., van der Klis, M. 1999, ApJ., 520, 262\par 
Remillard, R.A. 2005, Astron. Nachr., 326, 804 \par 
%Shafee, R., McClintock, J.E., Narayan, R., Davis, S.W., Li, L.-X.,
%Remillard, R.A. 2005, astro-ph/0508302\par     
Tosa, M. 1994, ApJ, 426, L81 \par
%Miller, M.C., Lamb, F.K., \& Psaltis, D. 1998, ApJ, 508,791\par
%van der Klis, M. 2000, ARA\&A, 38, 717    \par
van der Klis, M. 2004, in Compact stellar X-ray sources (Cambridge University Press), 
   eds. W.H.G. Lewin and M. van der Klis (astro-ph/0410551)    \par
%van der Klis, M., Wijnands, R. A. D., Horne, K., \& Chen, W. 1997, ApJ,
%     481, L97 \par
Whitehurst, R. 1988, MNRAS 232, 35    \par
\bigskip\bigskip

\end{document}